\documentclass[12pt,a4paper]{article}
\usepackage{amsfonts}
\usepackage{amsmath}

\input{tcilatex}

\begin{document}

\title{Line Laplace Transforms for obtaining the Exact Bound States for the
Morse Potential}
\author{S.--A. Yahiaoui\thanks{%
E-mail: s\_yahiaoui@mail.univ-blida.dz}, M. Bentaiba\thanks{%
Correspondin author *E-mail: bentaiba@mail.univ-blida.dz} \\
%EndAName
LPTHIRM, D\'{e}partement de Physique, Facult\'{e} des Sciences,\\
Universit\'{e} Saad DAHLAB de Blida, Blida, Algeria}
\maketitle

\begin{abstract}
The line Laplace transforms is applied to the Morse potential. The
wavefunctions and the energy levels through suitable path of integration are
drived.


\end{abstract}

\section{Introduction}

The Morse potential [1] is one of the simplest example of the Natanzon
potentials [2] which has a finite number of the bound--states. It was
introduced by P. M. Morse in 1929 as a model to describing the vibrational
energy of a diatomic molecule and takes the form%
\begin{equation}
V\left( x\right) =V_{0}\left( 1-\func{e}^{-ax}\right) ^{2},  \tag{1}
\end{equation}%
where $x$ represents a bond length, $a$ a parameter controlling the
potential width, and $V_{0}$ the dissociation energy of the molecule.

Several authors have investigated the exact solvability of Schr\"{o}dinger
equation with the Morse potential using various approaches and numerous
papers have been increased in recent years, dealing with supersymmetric
quantum mechanics [3,4,5], $\mathfrak{so}\left( 2,1\right) $ and $\mathfrak{%
su}\left( 1,1\right) $ Lie algebras [6,7,8,9], the point canonical
transformations [10], variational method [11], path--integarl approach [12],
coherent states [13] and recently the Nikiforov--Uvarov method [14], are
used in order to provide the exact solution of the eigenfunctions and
corresponding energy eigenvalues for the Morse potential. It is also well
known that the Morse potential have a causal connection with the \ P\"{o}%
schl--Teller potential [15] and Coulomb potential [16].

Perhaps the must useful integral transforms frequently used in mathematical
physics and physical applications is Laplace transforms [17,18]. The main
application of the Laplace transforms consists probably in converting
differential equation into simpler forms that may be solved easily [19,20].

In the present paper, we will perform the line Laplace transforms to solve
the Schr\"{o}dinger equation with the Morse potential. We investigate how
the mathematical formalism of this method works to deduce the wavefunctions
and energy levels for the Morse potential. This simply means that the choice
of the path of integration $l$ must be specified. We will see that unless
the integrand has special properties, that lead the wavefunction integral to
depend only on the value of the end points; the value will depend on the
particular choice of the contour $l$.

This work is organized as follows: in the second section we perform the line
Laplace transforms in order to deal with the Morse potential. Section three
is devoted to obtaining the exact bound--states for the Morse potential by
choosing a suitable path of integration and in the last section we do our
final conclusion.

\section{Condition on the wavefunction $F\left( \protect\xi \right) $}

Following [19] and [20], the Schr\"{o}dinger equation of the Morse potential
(1) is given by%
\begin{equation}
\left[ \frac{d^{2}}{dx^{2}}-\frac{2mV_{0}}{\hbar ^{2}}\func{e}^{-2ax}+\frac{%
4mV_{0}}{\hbar ^{2}}\func{e}^{-ax}+\frac{2m}{\hbar ^{2}}\left(
E-V_{0}\right) \right] \psi \left( x\right) =0.  \tag{2}
\end{equation}

Introducing the new change of variable and new wavefunction as%
\begin{eqnarray}
\xi &=&k\func{e}^{-ax},\quad \text{with\quad }k=\frac{2\sqrt{2mV_{0}}}{%
a\hbar },  \TCItag{3.a} \\
\psi \left( x\right) &=&\xi ^{\mu }F\left( \xi \right) ,  \TCItag{3.b}
\end{eqnarray}%
where $\xi \in \left[ 0,\infty \right] $ and $\mu $ is a constant, allow to
transforming Eq.(2) into%
\begin{equation}
\left[ \xi ^{2}\frac{d^{2}}{d\xi ^{2}}+\left( 2\mu +1\right) \xi \frac{d}{%
d\xi }-\frac{1}{4}\xi ^{2}+\frac{k}{2}\xi +\mu ^{2}-\beta ^{2}\right]
F\left( \xi \right) =0,  \tag{4}
\end{equation}%
where $\beta =\sqrt{-\frac{2m}{a^{2}\hbar ^{2}}\left( E-V_{0}\right) }$.
Putting $\mu =-\beta $ [19,20], we get the differential equation%
\begin{equation}
\left[ \xi \frac{d^{2}}{d\xi ^{2}}-\left( 2\beta -1\right) \frac{d}{d\xi }-%
\frac{1}{4}\xi +\frac{k}{2}\right] F\left( \xi \right) =0,  \tag{5}
\end{equation}%
where Eq.(5) accepts a regular point at $\xi =0$ and an irregular one at $%
\xi =\infty $.

Let us consider the function $F\left( \xi \right) $ can be expressed as an
integral of the general form%
\begin{equation}
F\left( \xi \right) =\doint\nolimits_{l}f\left( t\right) \func{e}^{\xi t}dt,
\tag{6}
\end{equation}%
where $f\left( t\right) $ is an unknown function and $l$ is the path of
integration which does not depend on $\xi $. The integral (6) is often
called the \textit{line Laplace transforms.}

Then, by applying the derivative of $F\left( \xi \right) $ with respect to $%
\xi $, one obtains%
\begin{eqnarray}
F^{\prime }\left( \xi \right) &=&\doint\nolimits_{l}t\text{ }f\left(
t\right) \func{e}^{\xi t}dt,  \TCItag{7.a} \\
F^{\prime \prime }\left( \xi \right) &=&\doint\nolimits_{l}t^{2}f\left(
t\right) \func{e}^{\xi t}dt.  \TCItag{7.b}
\end{eqnarray}

Multiplying Eqs.(6) and (7.b) by $\xi $ and performing derivation by parts,
we get%
\begin{eqnarray}
\xi F\left( \xi \right) &=&\left\{ f\left( t\right) \func{e}^{\xi t}\right\}
_{l}-\doint\nolimits_{l}\frac{df\left( t\right) }{dt}\func{e}^{\xi t}dt, 
\TCItag{8.a} \\
\xi F^{\prime \prime }\left( \xi \right) &=&\left\{ t^{2}f\left( t\right) 
\func{e}^{\xi t}\right\} _{l}-\doint\nolimits_{l}\frac{d\left[ t^{2}f\left(
t\right) \right] }{dt}\func{e}^{\xi t}dt,  \TCItag{8.b}
\end{eqnarray}%
where the symbol $\left\{ Y\left( t\right) \right\} _{l}$ denotes the
increase of $Y\left( t\right) $ when $t$ describes the contour $l$.

Substituting Eqs.(6), (7.a) and (8.b) into the differential equation (5), we
obtain%
\begin{equation}
\left[ f\left( t\right) \left( t^{2}-\frac{1}{4}\right) \func{e}^{\xi t}%
\right] _{l}-\doint\nolimits_{l}\left\{ \frac{d\left[ t^{2}f\left( t\right) %
\right] }{dt}-\frac{1}{4}\frac{df\left( t\right) }{dt}+\left[ \left( 2\beta
-1\right) t-\frac{k}{2}\right] f\left( t\right) \right\} \func{e}^{\xi
t}dt=0.  \tag{9}
\end{equation}

The contour of integration $l$ is then chosen so that the first term in
Eq.(9) vanishes and that the integrand will vanish at the end points; these
considerations lead to write%
\begin{equation}
\frac{d\left[ t^{2}f\left( t\right) \right] }{dt}-\frac{1}{4}\frac{df\left(
t\right) }{dt}+\left[ \left( 2\beta -1\right) t-\frac{k}{2}\right] f\left(
t\right) =0.  \tag{10}
\end{equation}

Taking into account the differentiation of Eq.(10) and integrating the
result, we get%
\begin{equation}
f\left( t\right) =\mathcal{C}_{p,q}\left( t-\frac{1}{2}\right) ^{p-1}\left(
t+\frac{1}{2}\right) ^{q-1},  \tag{11}
\end{equation}%
where $\mathcal{C}_{p,q}$ is a constant of integration and the parameters $p$
and $q$ are defined as%
\begin{equation}
p-1=\frac{k}{2}-\frac{2\beta +1}{2},\qquad q-1=-\frac{k}{2}-\frac{2\beta +1}{%
2},  \tag{12}
\end{equation}%
and then the integral (6) may be written%
\begin{equation}
F\left( \xi \right) =\mathcal{C}_{p,q}\text{ }\doint\nolimits_{l}\left( t-%
\frac{1}{2}\right) ^{p-1}\left( t+\frac{1}{2}\right) ^{q-1}\func{e}^{\xi
t}dt,  \tag{13}
\end{equation}%
is a solution of differential equation (5) and the contour, by virtue of the
condition quoted above, must verify%
\begin{equation}
\left\{ \left( t-\frac{1}{2}\right) ^{p}\left( t+\frac{1}{2}\right) ^{q}%
\func{e}^{\xi t}\right\} _{l}\equiv 0.  \tag{14}
\end{equation}

\section{Wavefunctions and energy levels}

From Eq.(12), the parameters $p$ and $q$ are not integer and defined positive%
\footnote{%
We verify this assumption below; cf. relationship (19.a--b).}, then the
integrand in (13) has a \textit{branch points} at $t_{\pm }=\pm \frac{1}{2}$
[17], and the product $\left( t-\frac{1}{2}\right) ^{p}\left( t+\frac{1}{2}%
\right) ^{q}$ in Eq.(14) will vanish for $t_{\pm }$. The integrand into (13)
is therefore \textit{single--Valued} for the contour encircling both branch
points; i.e. taking the line segment joining $t_{+}=\frac{1}{2}$ and $t_{-}=-%
\frac{1}{2}$ as a \textit{cut line }[17].

Let us consider a particular choice of the contour such as%
\begin{equation}
l=\left\{ t\in \left[ -\frac{1}{2},\frac{1}{2}\right] \ /s=\xi \left( t+%
\frac{1}{2}\right) \right\} ,  \tag{15}
\end{equation}%
therefore Eq.(13) becomes%
\begin{equation}
F\left( \xi \right) =\left( -1\right) ^{p-1}\mathcal{C}_{p,q}\func{e}^{-\xi
/2}\text{ }\xi ^{1-p-q}\dint\nolimits_{0}^{\xi }s^{q-1}\left( \xi -s\right)
^{p-1}\func{e}^{s}ds.  \tag{16}
\end{equation}

Using the integral representation [21, cf. \textbf{9.211} 2, pp. 1058]%
\begin{equation}
_{1}F_{1}\left( \alpha ;\gamma ;\xi \right) =\frac{1}{B\left( \alpha ,\gamma
-\alpha \right) }\xi ^{1-\gamma }\dint\nolimits_{0}^{\xi }s^{\alpha
-1}\left( \xi -s\right) ^{\gamma -\alpha -1}\func{e}^{s}ds,\quad \left[ 
\func{Re}\gamma >\func{Re}\alpha >0\right] ,  \tag{17}
\end{equation}%
the integral into Eq.(16) becomes%
\begin{equation}
\dint\nolimits_{0}^{\xi }s^{q-1}\left( \xi -s\right) ^{p-1}\func{e}%
^{s}ds=B\left( q,p\right) \text{ }_{1}F_{1}\left( 1-p;2-p-q;\xi \right) , 
\tag{18}
\end{equation}%
where $_{1}F_{1}\left( \alpha ;\gamma ;\xi \right) $ is the confluent
hypergeometric function and $B\left( \alpha ,\gamma -\alpha \right) =\frac{%
\Gamma \left( \alpha \right) \Gamma \left( \gamma -\alpha \right) }{\Gamma
\left( \gamma \right) }$ is the beta function [17], with $p+q=\gamma $ and $%
\alpha =q$. In terms of Eqs.(12), the parameters in Eq.(18) read%
\begin{eqnarray}
1-p &\equiv &1+\alpha -\gamma =-\frac{k}{2}+\frac{2\beta +1}{2}, 
\TCItag{19.a} \\
2-p-q &\equiv &2-\gamma =2\beta +1.  \TCItag{19.b}
\end{eqnarray}

Then, the function $F\left( \xi \right) $ reads as 
\begin{equation}
F\left( \xi \right) =\mathcal{N}_{p,q}\func{e}^{-\xi /2}\xi
^{1-p-q}{}_{1}F_{1}\left( 1-p;2-p-q;\xi \right) ,  \tag{20}
\end{equation}%
where $\mathcal{N}_{p,q}=\left( -1\right) ^{1-p}\mathcal{C}_{p,q}B\left(
q,p\right) $.

However, it is well known that the (confluent) hypergeometric function
becomes a simple polynomial if and only if the parameter $1-p$ in Eq.(20)
equals $0$ or a negative integer [17]. We limit ourselves here to second
case, i.e. $1-p=-n$, leading to identify%
\begin{equation}
k-(2\beta +1)=2n,  \tag{21}
\end{equation}%
where $n$ is called the vibrational quantum number and takes the values $%
n=0,1,2,\ldots n_{\max }$. Substituting Eqs.(19.a--b) and (21) into Eq.(20),
and the result into Eq.(3.b), the wavefunction $\psi \left( \xi \right) $ is
then given by%
\begin{eqnarray}
\psi \left( \xi \right) &=&\xi ^{\mu }F\left( \xi \right)  \notag \\
&=&\mathcal{N}_{p,q}^{(n)}\text{ }\func{e}^{-\xi /2}\text{ }\xi ^{\beta }%
\text{ }_{1}F_{1}\left( -n;2\beta +1;\xi \right) .  \TCItag{22}
\end{eqnarray}

Inserting the parameters $\beta $ and $k$ as defined hereabove into Eq.(21),
then the corresponding energy levels are%
\begin{equation}
E_{n}=-V_{0}\left[ 1-\frac{a\hbar }{\sqrt{2mV_{0}}}\left( n+\frac{1}{2}%
\right) \right] ^{2}+V_{0}.  \tag{23}
\end{equation}%
and which Eqs.(22) and (23) agree with those obtained by [19,20].

\section{Conclusion}

We investigated a simple method of line Laplace transforms of finding the
wavefunctions and the corresponding energy levels of the Schr\"{o}dinger
equation with the Morse potential. The approach developed here has not
brought anything new, however, the main purpose is to investigate how the
method of line Laplace transforms works. We have shown that applying the
specific choice of the contour $l$, we can find one of the various integral
representations of the confluent hypergeometric functions and which are
associated with wavefunctions of the Morse potential.

Since there are other closely integral representations of confluent
hypergeometric functions, we attempt that they can be other contours of
integration used to get the wavefunctions of the Morse potential.

\end{document}